\documentclass[twocolumn,showpacs,amsmath,aps]{revtex4}
\usepackage{graphicx,color}
\usepackage{CJK}
\usepackage{bm}
\usepackage[hypertex]{hyperref}

\newcommand{\be}{\begin{equation}}
\newcommand{\ee}{\end{equation}}
\newcommand{\bea}{\begin{eqnarray}}
\newcommand{\eea}{\end{eqnarray}}
\newcommand{\bsube}{\begin{subequations}}
\newcommand{\esube}{\end{subequations}}

\newcommand{\Eq}[1]{Eq.\,(\ref{#1})}

\newcommand{\dg}{\dagger}
\newcommand{\la}{\langle}
\newcommand{\ra}{\rangle}

\newcommand{\nl}{\nonumber \\}

\newcommand{\bsub}{\begin{subequations}}
\newcommand{\esub}{\end{subequations}}

\newcommand{\ket}[1]{{\left| #1 \right\rangle }}

\begin{document}
\begin{CJK*}{GBK}{Song}


\title{Weak values in continuous weak measurement of qubits}

\author{Lupei Qin}
\affiliation{Department of Physics, Beijing Normal University,
Beijing 100875, China}
\author{Pengfei Liang}
\affiliation{Department of Physics, Beijing Normal University,
Beijing 100875, China}
\author{Xin-Qi Li}
\email{lixinqi@bnu.edu.cn}
\affiliation{Department of Physics, Beijing Normal University,
Beijing 100875, China}

\date{\today}

\begin{abstract}
For continuous weak measurement of qubits,
we obtain exact expressions for weak values (WVs) from
the post-selection restricted average of measurement outputs,
by using both the quantum-trajectory-equation (QTE)
and quantum Bayesian approach.
The former is applicable to short-time weak measurement,
while the latter can relax the measurement strength to finite.
We find that even in the ``very" weak limit
the result can be essentially different from the one
originally proposed by Aharonov, Albert and Vaidman (AAV),
in a sense that our result
incorporates {\it non-perturbative}
correction which could be important when the AAV's WV is large.
Within the Bayesian framework, we obtain also elegant
expressions for {\it finite} measurement strength
and find that the amplifier's noise in quantum measurement
has no effect on the WVs.
In particular, we obtain very useful result for homodyne
measurement in circuit-QED system, which allows for
measuring the real and imaginary parts of the AAV's WV
by simply tuning the phase of the local oscillator.
This advantage can be exploited
as efficient state-tomography technique.
\end{abstract}


\maketitle

\section{Introduction}

The concept of weak value (WV)
was introduced by Aharonov, Albert and Vaidman (AAV)
in the seminal paper \cite{Aha88}
and was elaborated with extended discussions \cite{Ste89,Aha90}.
The most interesting and surprising feature of the WV is
that it can exceed the range of eigenvalues of the observable.
Obviously, the appearance of this nonclassical or strange WV
violates our common knowledge that the quantum average
must be bounded by the extremum of eigenvalues.
Therefore its interpretation has been a
subject of confusion \cite{Leg89,Aha01}.
Despite of this, the strange WV
has been experimentally confirmed \cite{Rit91,Wis05,DiC13}.

In addition to theoretical curiosity,
the concept of WV is very useful to facilitate
to resolve and generate quantum paradoxes,
such as the Hardy's paradox \cite{Aha02,Pay04},
the superluminal transport in optical fiber \cite{Bru0304}
or in vacuum \cite{Aha02a},
the momentum-disturbance relationship \cite{Wis07}
and locally averaged momentum streamlines \cite{Hi12,Stbg11}
in two-slit interferometer,
and the more recently proposed quantum Cheshire Cat
\cite{Aha13,Ban14,Has14}.
Even more, the WV has novel applications in technological aspects,
e.g., for weak signal amplification and sensitive estimation
of unknown small parameters \cite{Kwi08,How09},
and for efficient quantum state tomography
\cite{Lun11,Lun12,Sa12}.


In AAV's original work, the WV of spin-$\frac{1}{2}$ particle
was proposed as
$ \langle \sigma^z\rangle_w
={\rm Re}\frac{\langle \psi_f|\sigma^z| \psi_i\rangle}
{\langle \psi_f|\psi_i\rangle}
\equiv {\rm Re} (\sigma^z_w)$,
where $|\psi_i\rangle$ and $|\psi_f\rangle$
are the pre- and post-selected states (PPS).
Note that the key quantity (we term it in this work as AAV's WV),
$\sigma^z_w= \frac{\langle \psi_f|\sigma^z| \psi_i\rangle}
{\langle \psi_f|\psi_i\rangle}$, is a complex object.
Its interpretation is thus subtle
and has caused debates \cite{Ste89,Aha90,Leg89,Aha01}.
Simpler understanding for WV is from the
practical measurement point of view \cite{Wis05,DiC13},
which is to be termed as {\it measured} WV
or simply as WV in this work.
The measured WV corresponds to a PPS restricted average,
say, a sub-ensemble-average of the {\it weak measurement} results,
with partial data survived from the post-selection.


In most of the existing WV studies, the {\it weak measurement}
is implemented via a {\it unitary} interaction model
by two coupled subsystems,
i.e., the measured system and the measuring device,
while the device is described by a {\it quantum}
Gaussian probe wavefunction with continuous variables
which is subject to extra projective measurement
for the ``pointer" coordinate or momentum.
For this type of model and using
a conditional-probability formulation
the WV can be re-derived as \cite{Dre14}:
\bea\label{WV-1}
\langle \sigma^z\rangle_w &=& \frac{_f\langle x\rangle_i}{\epsilon}
= \int \frac{x}{\epsilon} P_x(f;i) dx  \nl
&=& {\rm Re}(\sigma^z_w )
+ O(\epsilon^2) ,
\eea
where $\epsilon$ is the unitary interaction strength
which determines also the average shift of the pointer
caused by the system eigenstate.
$P_x(f;i)$ is the success probability of post-selection
with $| \psi_f\rangle$, following the $x$-conditioned
system state after the weak measurement.
Note that this post-selection conditioned average is also
the method of data analysis in experiments \cite{Wis05,DiC13}.


One may note that, differing from the unitary interaction
model mentioned above for weak measurements,
there exists alternative class of models and real systems,
which has been extensively studied for
qubit measurement and control,
especially in solid-state setup using such as
the quantum point contact (QPC) \cite{SG97,Kor01,GM01},
the single electron transistor (SET) \cite{Sch98,Sch01},
or the homodyne detection in circuit quantum-electrodynamics (cQED)
\cite{Bla04,Wall04,Sid12,Dev13,Sid13}.
However, despite particular importance,
this type of weak measurement has not yet received much attention
in the context of WV studies. Its main difference from
the unitary interaction model is that the ``device" in this type
of measurement scheme is described ``classically",
without involving the usual quantum wavefunction description.
Instead, this type of weak measurement is characterized
by quantum trajectories ---
the stochastic evolutions of the system state
conditioned on the (continuous) measurement outputs,
which can be described by
quantum-trajectory-equation (QTE) or quantum Bayesian rule.
In Ref.\ \cite{Wis02}, applying QTE and
using the conditional-probability method,
the canonical AAV result of \Eq{WV-1}
was recovered for continuous weak measurement of qubit.


In present work, along the line of Ref.\ \cite{Wis02},
we recalculate the WVs for this type of weak measurement.
Beyond \Eq{WV-1}, we obtain the following better result:
\begin{equation} \label{WV-2}
    \frac{ _{f}\langle x \rangle_{i}}{\epsilon}
    =\frac{ {\rm Re}(\sigma^z_{w})}{1+g(|\sigma^z_{w}|^{2}-1)} \,.
\end{equation}
Similar as in \Eq{WV-1},
the scaling parameter $\epsilon$ in this expression
is the shift of the average output current
caused by the basis state of the qubit.
Notably, in the denominator,
$g$ characterizes the measurement strength
and takes a simple form $g=\gamma dt$ with
$\gamma$ the measurement rate and $dt$ the measurement time.

Remarkable modification of \Eq{WV-2} over \Eq{WV-1}
is the ``partial"-summation-type term in the denominator.
Actually, as long as the expansion $e^{-2g}\simeq 1-2g$
holds to be valid, \Eq{WV-2} is the exact result of the WV.
This conclusion can be proved by calculating
$_{f}\langle x \rangle_{i}$ via the quantum Bayesian rule,
which does not restrict the measurement strength to weak limit.
We will see that the exact WV result for arbitrary
measurement strength corresponds to a simple replacement
$g\rightarrow {\cal G}=(1-e^{-2g})/2$ in \Eq{WV-2},
while the measurement strength reads
$g=\gamma t_m$, with $t_m$ now a finite time of measurement.

The non-perturbative correction in \Eq{WV-2}
cannot be neglected when $\sigma^z_{w}$ is large
and should be taken into account in the various
applications of WVs.
For instance, in the context of WV amplifications,
it has been pointed out that some ``nonlinear" correction
to the linear result of AAV WV would set an upper bound
for the amplification coefficient, based on
similar results obtained within the framework of the
unitary interaction model of weak measurements
\cite{Wu11a,Wu11b,Tana11,Naka12,Sus12,Kof12}.
As a first and simple remark,
our present work differs from those in that
we consider different type of weak measurement, which
allows for employing different formulations to obtain
non-perturbative and exact WV expressions
for the measurement of qubits under consideration,
in particular obtaining useful result
for the solid-state superconducting circuit-QED setup.

The remaining part of this work is organized as follows.
In Sec.\ II, we complete all the formal aspects of calculating
the WVs for the continuous weak measurement of solid-state qubits.
In Sec.\ III we specify the study to solid-state
circuit-QED setup to carry out useful result which allows for
direct measurement of the real and imaginary parts of the AAV WV,
by simply tuning the local oscillator's phase in the homodyne measurement.
In Sec.\ IV we summarize the work with brief remarks.

\section{Weak values of qubit measurements}

\subsection{Continuous Measurement of Qubit}

There are several experimentally accessible systems
for continuous weak measurement of qubits, including for instance
the solid-state charge qubit measured by quantum-point-contact
or single-electron-transistor \cite{SG97,Kor01,GM01,Sch98,Sch01},
and the superconducting qubit in circuit-QED \cite{Bla04,Wall04}
measured via the cavity-field homodyne detection \cite{Sid12,Dev13,Sid13}.
Under reasonable simplifications,
all these setups can be commonly described as follows.
First, the qubit Hamiltonian is reduced to a two-state model
$H_q=\frac{\Delta}{2}\sigma^z$,
where $\sigma^z=|1\ra\la 1|-|2\ra\la 2|$ and $\Delta=E_1-E_2$.
Second, the measurement principle is based on the {\it distinct}
stationary current in the detector, $\bar{I}_{1,2}$,
which is associated with the qubit state $|1\ra$ or $|2\ra$.
Moreover, of particular interest is continuous weak measurement
for a superposition state $|\psi_i\ra=\alpha|1\ra +\beta|2\ra$.
In this case, the output current can be expressed in general as
$I(t)=I_0+\frac{\Delta I}{2}\la \sigma^z\ra + \xi_I(t)$,
where $I_0=(\bar{I}_1+\bar{I}_2)/2$ and $\Delta I=|\bar{I}_1-\bar{I}_2|$.
The last term obeys the statistics of white noise, with ensemble average
${\rm E}[\xi_I(t)\xi_I(t')]=\frac{S_0}{2}\delta(t-t')$,
where $S_0$ is the spectral density of the output current.
In this work, we would like to adopt a reduced description,
by denoting the measurement output as
$J(t)=\frac{I(t)-I_0}{\sqrt{S_0/2}}$.
Then we have
$J(t)=2\sqrt{\gamma}\la \sigma^z\ra+\xi(t)$,
where $\gamma=\frac{(\Delta I)^2}{8S_0}$
and ${\rm E}[\xi(t)\xi(t')]=\delta(t-t')$.

\subsubsection{QTE Approach}

Conditioned on the continuous outcomes of measurement given above,
the qubit state evolution (in the rotating frame with respect to $H_q$)
is governed by the following
It\^o-type quantum trajectory equation (QTE) \cite{GM01}
\bea\label{QTE-1a}
d\rho = \gamma \mathcal{D}[\sigma^{z}]\rho\,dt
  +\sqrt{\gamma} \mathcal{H}[\sigma^{z}]\rho\,dW(t) \,.
\eea
The super-operators are defined as
$\mathcal{D}[A]\rho=A\rho A^{\dagger}
-\frac{1}{2}\{A^{\dagger}A,\rho \}$,
and $\mathcal{H}[A]\rho = A\rho+\rho A^{\dg}
-\mathrm{Tr}[(A+A^{\dg})\rho]\rho$.
The Wiener increment is formally defined by $dW(t)=\xi(t)\,dt$,
in experiment which should be extracted from the output current.
Then, after a short-time ($dt$) measurement,
the state is updated as $\tilde{\rho}(t+dt)=\rho(t) + d\rho(t)$,
in the qubit basis which reads
\begin{subequations}\label{QTE-1b}
\begin{equation}\label{}
 \tilde{\rho}_{11}
 =\rho_{11}+2\sqrt{\gamma} (1-\langle \sigma^z\rangle)\rho_{11}\,dW \,,
\end{equation}
\begin{equation}\label{}
    \tilde{\rho}_{22}=\rho_{22}-2\sqrt{\gamma}
    (1+\langle\sigma^z\rangle)\rho_{22}\,dW   \,,
\end{equation}
\begin{equation}\label{}
   \tilde{\rho}_{12}=\rho_{12} -2\rho_{12}(\gamma\,dt
   +\sqrt{\gamma}\langle\sigma^z\rangle \,dW)  \,,
\end{equation}
\end{subequations}
where
$\langle\sigma^z\rangle={\rm Tr}[\sigma^z\rho(t)]=\rho_{11}-\rho_{22}$.
Below we compare this result with the one from the quantum Bayesian approach.

\subsubsection{Quantum-Bayesian Approach}

The quantum Bayesian approach, originally proposed by Korotkov \cite{Kor99},
is based on the well-known Bayes formula in Probability Theory
together with a {\it quantum purity} consideration.
The former is utilized to determine the diagonal elements
while the latter is for determination of the off-diagonal ones.
In quite compact form, the quantum Bayesian approach
updates the qubit state from $\rho$ to $\tilde{\rho}$
according to the following rule:
\bea\label{BR-1}
&&\tilde{\rho}_{11}=\rho_{11} P_{1}(x) / \mathcal{N}(x),
~~~\tilde{\rho}_{22}=\rho_{22} P_{2}(x)/ \mathcal{N}(x), \nl
&&
\tilde{\rho}_{12}=\rho_{12} \sqrt{P_{1}(x)P_{2}(x)} /\mathcal{N}(x) \,.
\eea
In these formulas, associated with the qubit state $|j\ra$ ($j=1,2$),
$P_{j}(x)$ is the {\it priori} probability distribution ``knowledge"
for the stochastic outputs ``$x$".
$\mathcal{N}(x)=\rho_{11}P_{1}(x)+\rho_{22}P_{2}(x)$
is a normalization factor.
Note that the last rule (for off-diagonal element)
is from a purity consideration.
About the distribution ``knowledge", in most cases,
the integrated output
(denoted here by the stochastic variable ``$x$")
satisfies a Gaussian statistics:
\begin{equation}\label{}
  P_{j}(x)=\frac{1}{\sqrt{2\pi D}}
  \exp\left[ - \frac{(x-\bar{x}_j)^{2}}{2 D}\right]  \,,
\end{equation}
where $\bar{x}_{1(2)}=\pm\epsilon$ denotes the average output
corresponding to qubit state $|1(2)\ra$,
and $D$ is the distribution variance.

As a more special case, let us consider a short-time measurement
over $dt$. The measurement result (stochastic ``$x$") simply
reads $x=J(t)dt=2\sqrt{\gamma}\langle \sigma^z \rangle\,dt+dW(t)$.
From this, we can identify
$\epsilon=2\sqrt{\gamma}dt$ and $D=dt$.
Viewing the small parameters $\epsilon$, $dt$ and $x$,
we first expand the exponential functions
in the Bayesian formulas to obtain
\begin{subequations}
\begin{equation}\label{}
    \tilde{\rho}_{11}=\rho_{11}\left(1+\frac{\epsilon x}{D}\right)
    \left(1+\frac{\epsilon x}{D}\langle\sigma^z\rangle\right)^{-1}
     \,,
\end{equation}
\begin{equation}\label{}
   \tilde{\rho}_{22}=\rho_{22}
   \left(1-\frac{\epsilon x}{D }\right)
    \left(1+\frac{\epsilon x}{D}\langle\sigma^z\rangle\right)^{-1}
   \,,
\end{equation}
\begin{equation}\label{}
    \tilde{\rho}_{12}=\rho_{12}\left[1+\frac{1}{4}
    (\frac{\epsilon x}{ 2D})^{4}\right]
    \left(1+\frac{\epsilon x}{D }\right)^{-1}
     \,.
\end{equation}
\end{subequations}
Performing further expansion up to the second-order
of $\epsilon$ and $x$, one can prove their equivalence
to \Eq{QTE-1a} or (\ref{QTE-1b}) as follows. Actually,
using the Bayesian expansion and applying the calculus
$\dot{\rho}_{ij}(t)=[\tilde{\rho}_{ij}(t+\frac{dt}{2})
- \tilde{\rho}_{ij}(t-\frac{dt}{2})]/dt$,
we obtain for instance
$ \dot{\rho}_{11}(t)=-\dot{\rho}_{22}(t)
= (2\epsilon x/ D) \rho_{11}(t)\rho_{22}(t)$,
which gives
\bea\label{QTE-1c}
\dot{\rho}_{11} = -\dot{\rho}_{22}
   =8\gamma\langle\sigma^z\rangle\rho_{11}\rho_{22}
   +4\sqrt{\gamma}\rho_{11}\rho_{22}\xi(t) \,.
\eea
Similarly, for the off-diagonal element, we have
\bea
\dot{\rho}_{12}
     =-4\gamma\langle\sigma^z\rangle^{2}\rho_{12}
     -2\sqrt{\gamma}\langle\sigma^z\rangle\rho_{12}\xi(t) \,.
\eea
We may find that these two equations differ from the QTE
(\ref{QTE-1a}) or (\ref{QTE-1b}).
Indeed, using the Bayesian rule in this manner,
we obtain the {\it Stratonovich-type} QTE,
but not the {\it It\^o}-type \cite{Kor01}.
To obtain the QTE of the It\^o-type, one needs to add
a term $({\cal F}/2)(d{\cal F}/d\rho_{ij})$ to each $\dot{\rho}_{ij}$,
where ${\cal F}$ is the factor before $\xi(t)$ in each equation \cite{GM01}.
Applying this conversion rule, one obtains
\begin{subequations}
\begin{equation}
    \dot{\rho}_{11} = - \dot{\rho}_{22}
    =4\sqrt{\gamma}\rho_{11}\rho_{22}\xi(t) \,,
\end{equation}
\begin{equation}\label{}
    \dot{\rho}_{12}=-2\gamma \rho_{12}
    -2\sqrt{\gamma}\langle\sigma^z\rangle\rho_{12}\xi(t) \,.
\end{equation}
\end{subequations}
This result is the same of \Eq{QTE-1a} or (\ref{QTE-1b}).

Below we make more comparisons.
Inserting the above Bayesian expansion into the It\^o calculus
$\dot{\rho}_{ij}(t)=[\tilde{\rho}_{ij}(t+dt)- \rho_{ij}(t)]/dt$,
one can obtain the It\^o-type QTE
under the {\it Milstein algorithm} \cite{KP},
for which we exemplify
\begin{eqnarray}
 d\rho_{11}=4\sqrt{\gamma} \rho_{11}\rho_{22}\, dW
   &-& 8\gamma\rho_{11}\rho_{22}(\rho_{11}-\rho_{22})  \nl
   && \times\, [(dW)^2-dt] \,.
\end{eqnarray}
Note that this result differs from
$d\rho_{11}=4\sqrt{\gamma} \rho_{11}\rho_{22} dW $,
which corresponds to the so-called {\it Euler algorithm} \cite{KP}.

Alternatively, if we only keep $dt$ and $dW$ as small parameters
in the Bayesian expansions and use the calculus
$\dot{\rho}_{ij}(t)=[\tilde{\rho}_{ij}(t+dt)- \rho_{ij}(t)]/dt$,
we find that expansion up to the first order of $dt$ and $dW$
corresponds to the Stratonovich-type QTE.
However, if keeping terms in the expansion up to $(dW)^2$,
one arrives to
the It\^o-type QTE under the {\it Milstein algorithm}.
This observation tells us that,
in order to make the result from the quantum Bayesian rule
be fully identical to that given by the It\^o-type QTE,
one should first make the expansion of the Baysian formula
up to $(d W)^2$, then set $(d W)^2=d t$.

\subsection{Weak Values}

\subsubsection{QTE Approach}

The post-selection restricted average can be calculated
using the following conditional-probability method:
\begin{equation}\label{Qfi}
_{f}\langle x \rangle_{i}=\frac{\int dx x
P_{i}(x) P_{x}(f)}{\int dx P_{i}(x) P_{x}(f)} \,
\end{equation}
where $x$ stands for the stochastic outcome of the weak measurement,
with probability $P_{i}(x)$ determined by the qubit state $|\psi_i\ra$.
Assuming
$|\psi_i\rangle=\alpha|1\rangle+\beta|2\rangle$,
we have $P_{i}(x)=|\alpha|^{2}P_{1}(x)+|\beta|^{2}P_{2}(x)$,
where $P_{1(2)}(x)$ is the priori probability
corresponding qubit state $|1(2)\ra$.
$P_{x}(f)$ is the success probability of the post-selection
by state $|\psi_f\ra$, following the measurement outcome $x$.
As explained above for the continuous measurement of qubit,
the output current can be expressed as
$J(t)=2\sqrt{\gamma}\langle \sigma^z\rangle+\xi(t)$.
For time interval $dt$, the output charge
(integrated current) is $x=J(t) dt$,
and the qubit state is updated via \Eq{QTE-1a} or (\ref{QTE-1b})
by inserting $dW = x-2\sqrt{\gamma}\langle \sigma^z\rangle dt$.
Then, the success probability of post-selection is
$P_{x}(f)=\langle \psi_f|\tilde{\rho}(t+dt)|\psi_f\rangle$.
More explicitly, we have
\bea
&& P_{x}(f)= P\,\left\{ 1+\gamma dt (|\sigma^z_{w}|^{2}-1) \right. \nl
& & \left. ~~~~~~~ +2\sqrt{\gamma}[{\rm Re}(\sigma^z_{w})
- \langle \sigma^z\rangle]dW \right\}  \,,
\eea
where $P=|\langle \psi_f|\psi_i\rangle|^{2}$
and $\sigma^z_w= \langle \psi_f|\sigma^z| \psi_i\rangle
/ \langle \psi_f|\psi_i\rangle$, as defined in Sec.\ I (Introduction).
Straightforwardly, we evaluate the numerator of \Eq{Qfi} as
\begin{eqnarray}
&& M_1 = P\{[1+4\gamma dt\langle\sigma^z\rangle({\rm Re}(\sigma^z_{w})
 -\langle\sigma^z\rangle)](|\alpha|^{2}\bar{x}_{1}
   +|\beta|^{2}\bar{x}_{2}) \nonumber\\
&& + 2\sqrt{\gamma}({\rm Re}(\sigma^z_{w})
-\langle \sigma^z\rangle)[|\alpha|^{2}(D+\bar{x}_{1}^{2})
   +|\beta|^{2}(D+\bar{x}_{2}^{2})] \}  \, ,  \nl
\end{eqnarray}
and the denominator as
\begin{eqnarray}
   M_2 &=& P[1+\gamma dt(|\sigma^z_{w}|^{2}-1)-4\gamma\langle\sigma^z\rangle
   dt({\rm Re}(\sigma^z_{w})-\langle \sigma^z\rangle)\nonumber\\
   && + 2\sqrt{\gamma}({\rm Re}(\sigma^z_{w})
   -\langle \sigma^z\rangle)(|\alpha|^{2}\bar{x}_{1}+|\beta|^{2}\bar{x}_{2})] \,.
\end{eqnarray}
Accounting for $|\alpha|^{2}+|\beta|^{2}=1$,
$|\alpha|^{2}-|\beta|^{2}=\langle\sigma^z\rangle$
and $\bar{x}_{1,2}=\pm 2\sqrt{\gamma}dt\equiv \pm\epsilon$,
the WV takes then a very compact form as
\begin{equation}\label{Qfi-2a}
    \frac{ _{f}\langle x\rangle_{i} }{\epsilon} = \frac
    {{\rm Re}(\sigma^z_{w})}{1+ g (|\sigma^z_{w}|^{2}-1)} \,.
\end{equation}
Here, we have denoted $g=\gamma dt$, which characterizes well
the measurement strength.
This result differs from the canonical AAV WV,
$_{f}\langle x\rangle_{i}/\epsilon = {\rm Re}(\sigma^z_{w})$,
by a ``partial"-summation-type correction in the denominator.
However, as we will see in the following, as long as the expansion
$e^{-2g}\simeq 1-2g$ holds, \Eq{Qfi-2a} is non-perturbative and exact.
As shown in Fig.\ 1,
even in the weak measurement limit (short time $dt$ or small $g$),
the role of this correction could be essential
when $\sigma^z_{w}$ is large. This implies that
in the various WV applications
such as weak signal amplification,
the non-perturbative correction in the denominator
of \Eq{Qfi-2a} should be taken into account
\cite{Wu11a,Wu11b,Tana11,Naka12,Sus12,Kof12}.

\begin{figure}
  \includegraphics[scale=0.35]{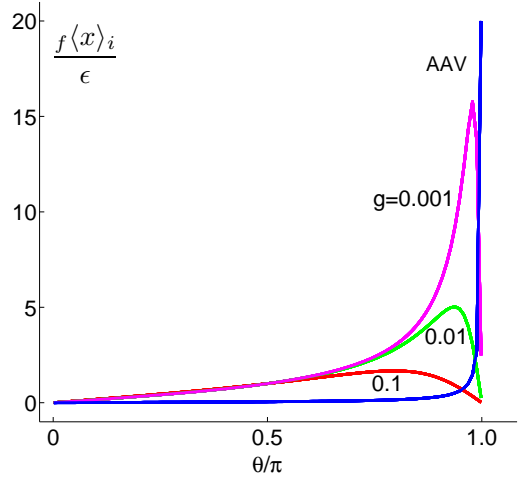}
\caption{(color online)
Non-perturbative correction of \Eq{Qfi-2a}
to the canonical AAV WV (blue line).
In this plot, we specify the initial state as
$|\psi_i\rangle=\frac{1}{\sqrt{2}}
(|1\rangle+|2\rangle)$
and the post-selection state as
$|\psi_f\rangle=[(\cos\frac{\theta}{2}+\sin\frac{\theta}{2})|1\rangle
+(\cos\frac{\theta}{2}-\sin\frac{\theta}{2})|2\rangle]/\sqrt{2}$,
which result in the AAV WV $\sigma^z_w=\tan(\theta/2)$.
We wee that, in contrast to the AAV WV,
a different trend appears in the large $\sigma^z_{w}$ regime. }
\end{figure}

\subsubsection{Bayesian Approach}

Applying the quantum Bayesian rule
we first calculate the post-selection probability,
$P_{x}(f)={\rm Tr} [\rho_{f}\tilde{\rho}(x)]$,
where $\rho_{f}=|\psi_f\rangle\langle \psi_f|$
is the post-selection state
and $\tilde{\rho}(x)$ the updated state from \Eq{BR-1}.
The result simply reads
\bea
    P_{x}(f)&=& \left[ \rho_{f11}\rho_{11}P_{1}(x)
    + \rho_{f22}\rho_{22}P_{2}(x)\right]  \nl
 & &+ 2{\rm Re}(\rho_{f12}^{*}\rho_{12})\sqrt{P_{1}(x)P_{2}(x)}  \,.
 \eea
$\rho_{ij}$ are the elements of the density matrix
of the initial state $\rho=|\psi_i\ra \la \psi_i|$,
which gives also the distribution function $P_i(x)$
in \Eq{Qfi} as $P_i(x)=\rho_{11}P_1(x)+\rho_{22}P_2(x)$.
Straightforwardly, after completing a number of Gaussian integrals,
the WV defined by \Eq{Qfi} is obtained as $_f\la x \ra_i=M_1/M_2$,
with $M_1$ and $M_2$ given by
\begin{subequations}\label{WV-BR}
\bea
 &&  M_1 = (\rho_{f11}\rho_{11}\bar{x}_{1}
   +\rho_{f22}\rho_{22}\bar{x}_{2} )
   +  (\bar{x}_{1}+\bar{x}_{2})
   \nl
 && ~~ \times {\rm Re}(\rho_{f12}^{*}\rho_{12})
   \exp\left[-(\bar{x}_{1}-\bar{x}_{2})^2 /(8D)\right] \,,
\eea
\bea
    M_2 &=& (\rho_{f11}\rho_{11}+\rho_{f22}\rho_{22})
    + 2\, {\rm Re}(\rho_{f12}^{*}\rho_{12}) \nl
&& \times \, \exp\left[-(\bar{x}_{1}-\bar{x}_{2})^2 /(8D)\right] \,.
\eea
\end{subequations}
Since the Bayesian rule works for finite strength measurement,
this result is then valid in general.

To be more specific, let us consider an initial state
$|\psi_i\rangle=\frac{1}{\sqrt{2}}
(|1\rangle+|2\rangle)$
and parameterize the post-selection state
$|\psi_f\rangle=d_1|1\rangle +d_2|2\rangle$
with polar angle $\theta$ as
$d_1=(\cos\frac{\theta}{2}+\sin\frac{\theta}{2})/\sqrt{2}$
and $d_2=(\cos\frac{\theta}{2}-\sin\frac{\theta}{2})/\sqrt{2}$.
Substituting the two states into \Eq{WV-BR}, we obtain
 \begin{equation}\label{WV-theta}
  _{f}\langle x\rangle_{i}=\frac{\bar{x}_{1}+\bar{x}_{2}}{2}
    + \frac{\sin\theta }
    {1+ G \cos\theta}
  \left(\frac{\bar{x}_{1}-\bar{x}_{2}}{2}\right)  \,,
 \end{equation}
where $G=\exp[-(\bar{x}_{1}-\bar{x}_{2})^2 /(8D)]$.
Similar as \Eq{WV-BR},
this result is valid for arbitrary strength of measurement.

Now consider the limit of weak measurement based on this result.
If we take first the limit
$G\rightarrow 1$, we find that
the WV can diverge when $\theta\rightarrow \pi$,
like the AAV WV, in present case
which reads $\sigma^z_w=\tan(\theta/2)$.
This limiting order corresponds also to
taking first the limit $g=\gamma dt\rightarrow 0$
in the denominator of \Eq{Qfi-2a},
then altering the post-selection to get divergent WV.
However, for any real measurement,
the measurement strength must be nonzero,
which results in $_{f}\langle x\rangle_{i}/\epsilon$
behaving as shown in Fig.\ 1.
The turnover behavior implies an existence of optimal
post-selection state which maximizes the WV,
and vanishing WV when the AAV WV diverges.
The latter corresponds to
$G\neq 1$ and $\theta\rightarrow \pi$ in \Eq{WV-theta}.

Moreover, it is desirable to recast the general WV result
\Eq{WV-BR} into more compact form,
like \Eq{Qfi-2a} in the weak measurement limit.
In the Bayesian result, let us identify
$\bar{x}_{1,2}=\pm 2\sqrt{\gamma}t_m=\pm \epsilon$ and $D=t_m$,
where $t_m$ is the finite measurement time.
Substituting these into \Eq{WV-BR}, after some algebra we get
\begin{equation}\label{WV-II}
    \frac{ _{f}\langle x\rangle_{i} }{\epsilon} = \frac
    {{\rm Re}(\sigma^z_{w})}{1+ {\cal G}\, (|\sigma^z_{w}|^{2}-1)} \,.
\end{equation}
Corresponding to $g=\gamma dt$ in \Eq{Qfi-2a},
the counterpart ${\cal G}$ in this result
has been generalized to ${\cal G}=(1-e^{-2g})/2$ with $g=\gamma t_m$.
Obviously, both results are identical for short time limit.
This proves also that the QTE result of WV is exact
in the weak measurement limit, having not lost anything
in the ``partial"-summation-type form of \Eq{Qfi-2a}.

\subsubsection{Noisy Amplification}

For many realistic measurements such as in circuit-QED
(to be addressed in Sec.\ III), the weak output signal
$x$ must be amplified properly.
As a simplified description, the principle of a linear amplifier
can be modeled by a second-tier of output voltage as
$v_x=v_{0}+x R$, where $R$ is the amplification
coefficient and $v_0$ denotes a reference voltage.
In addition to this one-to-one correspondence,
the amplifier will inevitably introduce extra noise.
Without loss of generality, we assume that the noise is Gaussian.
Then, the output voltage $\tilde{v}$ satisfies
$P_{x}(\tilde{v})= (2\pi\tilde{\sigma}^{2})^{-1/2}
\exp[- \frac{(\tilde{v}-v_x)^{2}}{2\tilde{\sigma}^{2}}]$.
Viewing the relation between $v_x$ and $x$,
we may reexpress this distribution alternatively as
\begin{equation}
    P_{x}(v)=\frac{1}{\sqrt{2\pi \sigma^{2}}}
    \exp\left[ -\frac{(v-x)^{2}}{2\sigma^{2}}  \right] \,,
\end{equation}
where we have re-defined $\sigma=\tilde{\sigma}/R$
and $v= (\tilde{v}-v_0)/R$.

Now we analyze the consequence of the amplifier's noise.
Let us consider first how the initial probability
of the first-tier output,
$P_i(x)=\rho_{11}P_{1}(x)+\rho_{22}P_{2}(x)$,
should be modified from the amplified signal
$\tilde{v}$ (or the re-scaled $v$).
This can be done by using the classical Bayes formula:
\bea
P_{v}(x)=\frac{P_i(x)P_{x}(v)}{\int dx P_i(x) P_{x}(v)} \,.
\eea
This distribution tells us that, for a given amplified
signal $v$, there are many possible $x$.
Therefore, the $v$-conditioned qubit state should be
\begin{equation}\label{}
    \tilde{\rho}_{v}=\int dx \tilde{\rho}_x P_{v}(x) \,,
\end{equation}
where $\tilde{\rho}_x$ is the qubit state conditioned on $x$,
in general which is given by the quantum Bayesian rule of \Eq{BR-1}.
More explicitly, we have
\begin{subequations}
\begin{equation}
    \tilde{\rho}_{v,11}=\rho_{11}{\cal B}_1(v)/{\cal N}_v  \,,
\end{equation}
\begin{equation}
    \tilde{\rho}_{v,22}=\rho_{22}{\cal B}_2(v)/{\cal N}_v  \,,
\end{equation}
\begin{equation}
    \tilde{\rho}_{v,12}=\rho_{12}{\cal B}_3(v)/{\cal N}_v  \,.
\end{equation}
\end{subequations}
Here, for the sake of brevity, we have introduced
${\cal B}_{1,2}(v)
= e^{-\frac{(v-\bar{x}_{1,2})^2}{2(D+\sigma^2)}}$,
${\cal B}_3(v)= e^{-\frac{(\Delta \bar{x})^2}{8D}}
e^{-\frac{(v-\bar{x}_0)^2}{2(D+\sigma^2)}}$
and
${\cal N}_v=\rho_{11}{\cal B}_1(v)+\rho_{22}{\cal B}_2(v)$,
where $\Delta \bar{x}=|\bar{x}_1-\bar{x}_2|$
and $\bar{x}_0=(\bar{x}_1+\bar{x}_2)/2$.

Similar as using \Eq{Qfi},
we can calculate now the ``noisy" WV by means of
\begin{equation}
    _{f}\langle x\rangle_{i}
    =\frac{\int dv\, v\, P_i(v) P_v(f)}
    {\int dv \, P_i(v) P_v(f)} \,,
\end{equation}
where $P_i(v)$ is the probability distribution of the
second-tier result ``$v$" which is given by
\bea
P_i(v)=\int dx P_i(x)P_x(v)
= \frac{\rho_{11}{\cal B}_1(v)+\rho_{22}{\cal B}_2(v)}
{\sqrt{2\pi(\tilde{\sigma}^2+DR^2)}} \,,
\eea
and the post-selection probability reads
\bea
P_v(f)=\rho_{f11}\tilde{\rho}_{v,11}
+\rho_{f22}\tilde{\rho}_{v,22}
+ 2\,{\rm Re}(\rho_{f12}\tilde{\rho}_{v,21}) \,.
\eea
In this way we obtain the result of WV defined from
the ``polluted" data, after noisy amplification.
However, being of some surprise is that we find
the same result of \Eq{WV-BR}.
This means that the WV is free from the amplifier's noise.
We may understand this result by the fact that
the noisy amplification of the second-tier
does affect the distribution width of the outputs,
but does not affect their average
even in the presence of post-selection.

The extra noise introduced in the amplification process is a sort
of non-ideality to reduce the quantum efficiency of measurement.
One may expect and actually can prove that other sources
of non-ideality owing to quantum information loss
does not affect the WV as well.
Desirably, the feature that the WV is free from amplifier's noise
and other sources of non-ideality (quantum information loss)
can benefit the measurement and applications of the WVs.

\section{Weak values in Circuit-QED}

In this section we specify the WV study carried out above
to the solid-state circuit-QED (cQED) system for two reasons.
First, the cQED system is one of the most promising
solid-state architectures for quantum information processing
\cite{Bla04,Wall04} and excellent platform for quantum
measurement and control studies \cite{Sid12,Dev13,Sid13}.
Second, this is to-date the most experimentally accessible
solid-state system where the continuous weak measurement
of the type considered in Sec.\ II has been realized
\cite{Sid12,Dev13,Sid13}.
In particular, we will obtain even more useful result than \Eq{Qfi-2a}
in that it allows for direct measurement of the full AAV WV, thus
for efficient method of qubit state tomography in this important system.

\subsection{QTE Approach}

Under reasonable approximations, the cQED system is well
described by the Jaynes-Cummings Hamiltonian \cite{Bla04}.
Moreover, in dispersive regime, by performing
a qubit-state-dependent displacement transformation
(called also ``polaron" transformation),
it is possible to eliminate the cavity degrees of freedom
to get a transformed QTE for the qubit state alone
under continuous homodyne measurements \cite{Gam08}:
\begin{eqnarray}\label{pQTE}
 &&  \dot{\rho} = -i\frac{\tilde{\omega}_q+B(t)}{2}[\sigma^z,\rho]
   +\frac{\Gamma_d(t)}{2}\mathcal{D}[\sigma^z]\rho       \nonumber \\
 &&  -\sqrt{\Gamma_{ci}(t)} \mathcal{M}[\sigma^z]\rho\xi(t)
   +i\frac{\sqrt{\Gamma_{ba}(t)}}{2}[\sigma^z,\rho]\xi(t)\,. \nl
 \end{eqnarray}
Here, $\tilde{\omega}_q=\omega_q+\chi$ is the renormalized
qubit energy owing to a dispersive shift $\chi$
to the bare energy $\omega_q$;
$B(t)=2\chi\mathrm{Re}[\alpha_1(t)\alpha^*_2(t)]$
is a generalized dynamic ac-Stark shift,
with $\alpha_1(t)$ and $\alpha_2(t)$
the cavity fields associated with the qubit states
$|1\ra$ and $|2\ra$, respectively.
In addition to the Lindblad term,
a new superoperator is introduced here by
$\mathcal{M}[\sigma^z]\rho \equiv (\sigma^z\rho+\rho\sigma^z)/2
-\la \sigma^z\ra \rho$,
with $\la \sigma^z\ra={\rm Tr}[\sigma^z\rho]$.
Also,
\begin{eqnarray}\label{GM3t}
  &&  \Gamma_d(t)=2\chi\mathrm{Im}[\alpha_1(t)\alpha^*_2(t)] \,,
  \nonumber \\
  && \Gamma_{ci}(t) = \kappa|\beta(t)|^2\cos^2(\varphi-\theta_\beta) \,,
  \nonumber \\
  &&  \Gamma_{ba}(t) = \kappa|\beta(t)|^2\sin^2(\varphi-\theta_\beta)\,,
\end{eqnarray}
characterize, respectively, the ensemble-average dephasing,
information-gain, and back-action rates.
In these expressions, we have introduced
$\beta(t)=\alpha_2(t)-\alpha_1(t)
\equiv |\beta(t)|e^{i\theta_{\beta}}$,
and denoted the cavity damping rate by $\kappa$
and the local oscillator (LO) phase by $\varphi$
for the homodyne measurement.

Within the framework of ``polaron" transformation,
the homodyne current of the cavity-field-quadrature
measurement is reduced to
$J(t)=-\sqrt{\Gamma_{ci}}\langle \sigma^{z}\rangle+\xi(t)$.
As explained in Sec.\ II,
from this result we can extract the Wiener increment via
$dW=x+\sqrt{\Gamma_{ci}}\langle \sigma^{z}\rangle dt$,
where $x= J(t)dt$ is the output of quadrature measurement.
Accordingly, we calculate the post-selection restricted
average of ``$x$" (the weak value) as follows.
First, with the help of the above QTE, we update the qubit state
from the ``initial" state $\rho(t)$,
which is denoted also by
$\rho_i=|\psi_i\rangle\langle \psi_i|$,
to $\tilde{\rho}(t+dt)$ based on the measurement result ``$x$".
Then, we calculate the success probability of post-selection via
$P_x(f)= \langle \psi_f|\tilde{\rho}(t+dt)|\psi_f\rangle$ which yields
\bea
 &&   P_x(f) = P \left\{ 1+ \left[ (\tilde{\omega}_q+B)
    {\rm Im}(\sigma^z_w)+\frac{\Gamma_{d}}{2}
    (|\sigma^z_w|^{2}-1)  \right] dt    \right.  \nl
&& ~~ \left. -\left[ \sqrt{\Gamma_{ci}} ({\rm Re}(\sigma^z_w)
    -\langle \sigma^z\rangle) + \sqrt{\Gamma_{ba}}
    {\rm Im}(\sigma^z_w)\right] dW  \right\} ,
\eea
where $P=|\langle \psi_f|\psi_i\rangle|^{2}$ and
$\sigma^z_w= \frac{\langle \psi_f|\sigma^z| \psi_i\rangle}
{\langle \psi_f|\psi_i\rangle}$,
as introduced above.
Further, replacing $dW$ with
$x+\sqrt{\Gamma_{ci}}\langle \sigma^{z}\rangle dt$
and carrying out the integration of
$\int dx x P(x)P_x(f)$ and $\int dx P(x)P_x(f)$, we obtain
\begin{equation}\label{WV1-cQED}
     _{f}\langle x\rangle_{i}
     = - \, \frac{ \epsilon_1\, {\rm Re}(\sigma^z_{w})
     +    \epsilon_2\, {\rm Im}(\sigma^z_{w}) }
     {1+[\, \tilde{\Omega}\, {\rm Im}(\sigma^z_w)
     +(\Gamma_d/2) (|\sigma^z_{w}|^{2}-1)\,]\,dt} \,.
\end{equation}
where $\epsilon_1=\sqrt{\Gamma_{ci}}\,dt$,
$\epsilon_2=\sqrt{\Gamma_{ba}}\,dt$ and $\tilde{\Omega}=\omega_q+B$.
In deriving this result we have identified
$\bar{x}_{1,2}=\mp \sqrt{\Gamma_{ci}}dt$
and $D=dt$, similar as in the previous section.

\Eq{WV1-cQED} is a useful result.
The particular structure in the numerator, i.e., the presence
of both the real and imaginary parts of $\sigma^z_w$,
allows for convenient measurement of the full complex AAV WV.
Indeed, from the expressions of $\Gamma_{ci}$ and $\Gamma_{ba}$,
we can selectively make either $\Gamma_{ci}$
or $\Gamma_{ba}$ be zero, by tuning the LO's phase $\varphi$.
Then, in short-time limit (which in most cases makes the denominator
near unity), one can separately measure
the real and imaginary parts of the AAV WV ($\sigma^z_w$).
A little bit complexity is for {\it unknown} initial state $|\psi_i\ra$.
In this case, the post-selection of state $|\psi_f\ra$
cannot rule out resulting in a large $\sigma^z_{w}$,
which would make the denominator of \Eq{WV1-cQED}
considerably deviate from unity.
In practice, one can solve this problem by iteratively
substituting the imprecise $\sigma^z_{w}$ into the
denominator of \Eq{WV1-cQED}
to obtain better estimation for $\sigma^z_{w}$.
Convergence is expected by a few such iterations.

The ability of measuring the real and imaginary parts of the complex
AAV's WV has important applications. One of them is to develop
new technique of state tomography \cite{Bam11,Bam12,Boy13}.
For such purpose in cQED system, one should identify
the ac-Stark shift and all the rates
($\Gamma_{ci}$, $\Gamma_{ba}$ and $\Gamma_d$) in \Eq{WV1-cQED}.
From \Eq{GM3t} we see that in general these quantities
are of time dependence, i.e.,
depending on the temporal evolution of the cavity field.
In experiments, however, the cQED setup is usually prepared in the
``bad-cavity" and weak-response limits \cite{Sid12,Dev13,Sid13,Kor11}.
In this case, the cavity-field would evolve to stationary state
on timescale considerably shorter than the quadrature
measurement time (denoted in this work by ``$dt$") \cite{Li14}.
Therefore, one can simply obtain the ac-Stark shift and all the rates
in \Eq{WV1-cQED} by using the {\it stationary} coherent-state fields
$\bar{\alpha}_1$ and $\bar{\alpha}_2$, which read
\bea\label{coh-state}
\bar{\alpha}_{1(2)}=-i\epsilon_m/[-i(\Delta_r\pm\chi)+\kappa/2],
\eea
where $\Delta_r=\omega_m-\omega_r$ is the offset
of the measurement and cavity frequencies.
For instance, in the bad-cavity and weak-response limit,
we obtain the stationary $B(t)$ as $B\simeq 2\chi\bar{n}$,
which recovers the standard ac-Stark shift
by noting that $\bar{n}=|\bar{\alpha}|^2$
and $\bar{\alpha}= -i\epsilon_m/(\frac{\kappa}{2})$.
Also, for resonant drive ($\omega_m=\omega_r$),
we have $\theta_{\beta}=0$.
Therefore one can measure the real and imaginary parts
of the AAV WV by choosing $\varphi=0$ and $\pi/2$, respectively.

\subsection{Bayesian Approach}

Let us continue to consider the WV for finite strength
measurement in cQED, applying the Bayesian scheme
proposed in Refs.\ \cite{Kor11,Li14}.
In general, for finite time ($t_m$) quadrature measurement,
the output corresponds to $x=x(t_m)=\int^{t_m}_{0}dt J(t)$,
where $J(t)=-\sqrt{\Gamma_{ci}}\langle \sigma^{z}\rangle+\xi(t)$.
Associated with qubit state $\ket{j}$,
the priori knowledge of distribution
$P_{j}(x)$ of the integrated quadrature $x$ is Gaussian,
given by $P_{j}(x)=\frac{1}{\sqrt{2\pi D}}
\exp[-(x-\bar{x}_j)^2/(2D)]$
where $\bar{x}_j=(-1)^j\sqrt{\Gamma_{ci}}\, t_m$ and $D=t_m$.

With the knowledge of $P_{j}(x)$,
the quantum Bayesian rule is the same as \Eq{BR-1}
for the diagonal elements, however for the off-diagonal element
it needs the following essential corrections \cite{Li14}
\begin{eqnarray}\label{BR-cQED}
&& \tilde{\rho}'_{12}(t_m)= \tilde{\rho}_{12}(t_m)
     ~ | \la \alpha_2(t_m)| \alpha_1(t_m) \ra |  \nl
&&  ~~~~~~~~~  \times   \exp\{-i[\Phi_1(t_m) + \Phi_2(t_m)]\} \,.
\end{eqnarray}
In this expression,
$\tilde{\rho}_{12}(t_m)$ is the updated result given by \Eq{BR-1}.
The overlap of the cavity-field coherent states,
$| \la \alpha_2(t_m)| \alpha_1(t_m) \ra |$,
characterizes the effect of purity degradation.
Other corrections are two phase factors:
\begin{subequations}\label{Phi12}
\bea
\Phi_1(t_m) = \int_0^{t_m} B(t) \;dt \,,
\eea
\bea
\Phi_2(t_m) = -\int_0^{t_m} \sqrt{\Gamma_{ba}(t)}\;J(t) \;dt \,.
\eea
\end{subequations}
The first factor stems from the energy shift of
the qubit owing to dynamic ac-Stark effect, and
the second one is from the last term of \Eq{pQTE}.
A tricky issue involved here may need to be reminded.
If we only consider
the last term of \Eq{pQTE}, the correction should be
$\Phi_2(t_m) = -\int_0^{t_m} \sqrt{\Gamma_{ba}(t)}\;\xi(t) \;dt $.
However, for each step of evolution over $(t,t+dt)$,
the stochastic output current is $J(t)$ which has a mean of
$\bar{J}(t)=-\sqrt{\Gamma_{ci}(t)}\langle \sigma^{z}\rangle_t$.
To guarantee the ensemble average of the off-diagonal element
over the stochastic $x=J(t)dt$
to be valid within the Bayesian scheme, one should adopt
$J(t)$ rather than $\xi(t)$ in the integrand of $\Phi_2(t_m)$.
This is consistent with \Eq{pQTE} after ensemble-average
over $\xi(t)dt=dW(t)$, in that the last term vanishes
and does not leave any phase factor such as
$e^{-i\sqrt{\Gamma_{ba}(t)}\bar{J}(t)dt}$ .

Similar to the short-time measurement analysis,
below we restrict our consideration
in the ``bad"-cavity and weak-response limit.
In this case, the ac-Stark field and all the rates are
independent of time, being given by the steady-state
cavity fields of \Eq{coh-state}.
As a consequence, the both phase factors are simplified to
$\Phi_1(t_m)=B\, t_m$ and $\Phi_2(t_m)=-\sqrt{\Gamma_{ci}}\,x(t_m)$,
and the purity-degradation factor
$| \la \alpha_2(t_m)| \alpha_1(t_m) \ra |$ can be approximated to unity.
After these,
the WV can be straightforwardly calculated using \Eq{Qfi},
where the post-selection probability $P_x(f)$ depends on the
updated state $\tilde{\rho}(t_m)$ from the cQED Bayesian rule,
particularly with \Eq{BR-cQED}.

Like \Eq{WV-BR}, for cQED we obtain $ _f\la x\ra_i = M_1/M_2$,
with $M_1$ and $M_2$ given by
\begin{subequations}\label{cQED-WV-2}
\bea
 && M_1 = -\sqrt{\Gamma_{ci}}\,t_m
  (\rho_{f11}\rho_{11} -\rho_{f22}\rho_{22})  \nl
 && ~~ + 2 \sqrt{\Gamma_{ba}}\,t_m e^{-\Gamma_d t_m}
  \,{\rm Im}( \rho_{f12} \rho^{*}_{12} e^{i\tilde{\Omega}t_m} )  \,,
\eea
\bea
&& M_2 = (\rho_{f11}\rho_{11}+\rho_{f22}\rho_{22}) \nl
&& ~~~~  + 2\,e^{-\Gamma_d t_m} {\rm Re}(\rho_{f12}\rho^*_{12}
e^{i\tilde{\Omega}t_m}) \,,
\eea
\end{subequations}
where $\Gamma_d=(\Gamma_{ci}+\Gamma_{ba})/2$
is the overall measurement-induced decoherence rate.
From this result, one can easily recover the WV of \Eq{WV1-cQED}
for short-time limit.
However, even for general finite time measurement,
we can obtain similar elegant expression as well.
After some algebra based on \Eq{cQED-WV-2}, we get
\bea\label{cQED-WV-3}
_f\la x\ra_i = - \frac
{\epsilon_1 {\rm Re}(\tilde{\sigma}^z_w)
+ \epsilon_2 {\rm Im}(\tilde{\sigma}^z_w)}
{1+{\cal G}\, (|\tilde{\sigma}^z_w|^2-1)} \,,
\eea
where $\epsilon_1 = \sqrt{\Gamma_{ci}}\,t_m $,
$\epsilon_2 = \sqrt{\Gamma_{ba}}\,t_m e^{-\Gamma_d t_m}$,
and ${\cal G}=(1-e^{-\Gamma_d t_m})/2$.
Notably, $\tilde{\sigma}^z_w$ in this result is
a slightly modified  AAV WV, taking a form as
$\tilde{\sigma}^z_w = \frac{\la\psi_f|\sigma_z|\tilde{\psi}_i\ra}
{\la\psi_f|\tilde{\psi}_i\ra}$,
where $|\tilde{\psi}_i\ra$ differs from the initial state
$|\psi_i\ra=c_1|1\ra + c_2|2\ra$ by a phase factor in terms of
$|\tilde{\psi}_i\ra = c_1 e^{-i\tilde{\Omega}t_m}|1\ra + c_2|2\ra$.

As a final remark, similar to the weak measurement limit
discussed above in previous subsection,
by tuning the LO phase $\varphi$ based on \Eq{cQED-WV-3},
one can still conveniently measure the real and imaginary parts of
$\tilde{\sigma}^z_w$, from which efficient state-tomography
technique can be developed for finite strength measurement
in cQED system. In practice this will make the experiment easier
since it allows for larger output signal.

\section{Conclusions}

To summarize, by applying both the quantum-trajectory-equation (QTE)
and quantum Bayesian approach,
we obtained non-perturbative and exact weak values (WVs)
for {\it continuous} weak measurement of qubits.
Differing from the usual unitary interaction model
for weak measurement in WV studies, the continuous
weak measurement scheme allows for non-perturbative
Bayesian approach to update the qubit state,
yielding thus exact WV result.
From this, by reducing the measurement strength (time duration),
we obtained exact expression for WVs in the weak measurement limit,
which is in full agreement with that derived by using the QTE scheme.

In particular, we have also extended the study to circuit-QED system.
The obtained desirable results allow for convenient
measurement of the real and imaginary parts of the AAV WV,
and thus for developing efficient technique of state-tomography.
Note that the cQED is to-date the most experimentally
accessible solid-state system where the continuous weak
measurement considered in this work has been realized.
Moreover, as analyzed in Sec.\ II B 3,
the WV measurement is free from the amplifier's noise
and/or the quantum efficiency of measurement.
Therefore, the cQED system is expected to be an ideal platform
for WV studies and related applications,
such as developing technique of direct state-tomography.

\vspace{0.1cm}
{\flushleft \it Acknowledgments.}---
This work was supported by the NNSF of China
under No.\ 91321106 and the State ``973" Project
under Nos.\ 2011CB808502 \& 2012CB932704.

\end{CJK*}
\end{document}